\documentclass[12pt]{iopart}

\usepackage{graphicx}
\usepackage{bm}
\usepackage{comment}

\begin{document}

\title[]
{Non-quasistatic response coefficients and dissipated availability for macroscopic thermodynamic systems}

\author{Yuki Izumida}

\address{Department of Complexity Science and Engineering, Graduate School of Frontier Sciences, The University of Tokyo, Kashiwa 277-8561, Japan}
\ead{izumida@k.u-tokyo.ac.jp}
\vspace{10pt}

\begin{abstract}
The characterization of finite-time thermodynamic processes is of crucial importance for extending equilibrium thermodynamics to nonequilibrium thermodynamics. 
The central issue is to quantify responses of thermodynamic variables and irreversible dissipation associated with non-quasistatic changes of thermodynamic forces applied to the system.
In this study, we derive a simple formula that incorporates the non-quasistatic response coefficients with Onsager's kinetic coefficients,
where the Onsager coefficients characterize the relaxation dynamics of fluctuation of extensive thermodynamic variables of semi-macroscopic systems. 
Moreover, the thermodynamic length and the dissipated availability that quantifies the efficiency of irreversible thermodynamic processes are formulated in terms of the derived non-quasistatic response coefficients.
The present results are demonstrated by using an ideal gas model.
The present results are, in principle, verifiable through experiments and are thus
expected to provide a guiding principle for the nonequilibrium control of macroscopic thermodynamic systems.
\end{abstract}

%
%
%
%
%

\section{Introduction}
In recent years, adiabatic control and even its shortcuts have received renewed and considerable interest both in theoretical and experimental points of view~\cite{DB2020,GRKTMM2019}.
In an adiabatic control, the parameters of a classical or a quantum system are sufficiently slowly varied compared to a system's relaxation time.
A quasistatic thermodynamic control is such a control and is one of the important building blocks of thermodynamics.
Consider a response of thermodynamic variables $\mathbf X$ upon quasistatically added small perturbation $\mathbf F$: 
\begin{eqnarray}
\mathbf x=\chi \mathbf F,\label{eq.response_ad}
\end{eqnarray}
where $\mathbf x$ is the deviation of $\mathbf X$ from the equilibrium value, and $\chi$ denotes the quasistatic response coefficient to the perturbation $\mathbf F$. 
According to equilibrium statistical mechanics, the quasistatic response coefficient $\chi$ can be calculated based on the equilibrium correlation function of fluctuations of thermodynamic variables in the absence of perturbation, which is referred to as the fluctuation-response relation~\cite{KTH1991,MPRV2008,YO}.

When the small perturbation is added slowly in time but non-quasistatically, eq.~(\ref{eq.response_ad}) may be generalized as
\begin{eqnarray}
\mathbf x=\chi \mathbf F-R\dot{\mathbf F},\label{eq.response_nad}
\end{eqnarray}
where the dot denotes the time derivative, and we call $R$ the non-quasistatic response coefficients.
A similar quantity to $R$ has been studied extensively, sometimes referred to as the friction tensor~\cite{SC2012} or the generalized friction coefficient~\cite{PP2021}. In the same manner as $\chi$, $R$ can be expressed in terms of temporal equilibrium correlation functions of fluctuations based on the linear response theory under an additional assumption of time-scale separation~\cite{SC2012,PP2021}.
The idea of time-scale separation has also been used in the context of adiabatic perturbation theory for quantum systems~\cite{ROP2008,SNBGD2022_1,SNBGD2022_2,SMDB2022}.

Remarkably, $R$ has been used as a metric tensor to define the thermodynamic length on a thermodynamic control space~\cite{SC2012,PP2021}. 
Such an introduction of geometric structure to thermodynamics went back to~\cite{W1975,R1979,R1995}, where the 
second derivative of internal energy~\cite{W1975} or entropy~\cite{R1979,R1995,SNI1984} with respect to extensive variables serves as a metric tensor on the thermodynamic space.
It was revealed that the thermodynamic length is closely related to the minimum dissipated availability~\cite{SB1983}.

The dissipated availability (or the entropy production rate) quantifies the efficiency of irreversible thermodynamic processes carried out by controlling parameters of a system in finite time~\cite{SB1983}.
It was originally proposed in~\cite{SB1983} for macroscopic systems based on an endoreversible assumption where a system is regarded in a local equilibrium state when perturbed non-quasistatically, but not in a global equilibrium with a heat reservoir (see also~\cite{FAQ1985,FAQS1985}).
The thermodynamic length and the dissipated availability have recently been formulated with an information-theoretic interpretation~\cite{C2007} for stochastic thermodynamic systems~\cite{DB2020,SC2012,PP2021,SS1997,ZSCD2012,KA1997,BD2014,M2015,KS2010}, which brought important applications to stochastic heat engine cycles~\cite{BS2020,MM2020,FD2022,FD2022_2,EB2022,YI2021,YI2022,WM2022,AL2020,MMPG2021}.

In this paper, we derive the non-quasistatic response coefficients of extensive thermodynamic variables for macroscopic thermodynamic systems against external thermodynamic forces, which are intensive.
Through the application of a singular perturbation theory~\cite{Stz} to the dynamics of fluctuations of extensive thermodynamic variables in the presence of 
time-dependent external thermodynamic forces, 
the non-quasistatic response coefficients in terms of Onsager's kinetic coefficients are derived. 
The Onsager's kinetic coefficients govern the relaxation dynamics of fluctuations of extensive variables of semi-macroscopic systems in the vicinity of the equilibrium state without the external thermodynamic forces.
Moreover, the derived non-quasistatic response coefficients are used for the formulation of the thermodynamic length and the dissipated availability. 
Because our formulation is based on the universal fluctuation dynamics of semi-macroscopic variables, it provides a more solid basis for the original formulation based on the endoreversible assumption~\cite{SB1983}.
Our results thus provide a guiding principle for the nonequilibrium control of macroscopic thermodynamic systems and contribute to the fundamental understanding of nonequilibrium thermodynamics.

\section{Setup}\label{Setup}
Consider a semi-macroscopic thermodynamic system.
Let $\hat{\mathbf X}=(\hat X_1, \cdots, \hat X_n)^{\rm T}$ ($n\ge 2$) be independent extensive thermodynamic variables of the system, where $\hat X_1=\hat U$ and $\hat X_2=\hat V$ are the internal energy and the volume by taking an entropy representation~\cite{SNI1984,C1985}. 
Here, ${\rm T}$ denotes the transpose, and the quantities with a hat denote random variables. 
The system is a small partial system of a large bath, where the system exchanges its extensive variables with the bath.
The bath is assumed to be sufficiently large so that its intensive parameters do not change upon exchanges of the extensive variables with the system.

Let $\hat{\mathbf x}=\delta \hat{\mathbf X}\equiv \hat{\mathbf X}-\mathbf X_{\rm eq}$ be the fluctuation of $\hat{\mathbf X}$,
where $\mathbf X_{\rm eq}$ is the equilibrium value of $\hat{\mathbf X}$.
At equilibrium, the spontaneous fluctuation $\hat{\mathbf x}$ obeys the celebrated Einstein's fluctuation formula~\cite{YO,LL}:
\begin{eqnarray}
\mathcal P_{\rm eq}(\hat{\mathbf x})\propto e^{\delta^2 S(\hat{\mathbf x})/k_{\mathrm B}},\label{eq.Einstein}
\end{eqnarray}
where $\mathcal P_{\rm eq}(\hat{\mathbf x})$ is the equilibrium probability distribution of $\hat{\mathbf x}$, $k_{\mathrm B}$ is the Boltzmann constant, and $\delta^2 S(\hat{\mathbf x})$ is 
the second-order entropy variation of the system from the equilibrium value serving as a potential function of the fluctuation $\hat{\mathbf x}$:
\begin{eqnarray}
\delta^2 S(\hat{\mathbf x})=-\frac{1}{2}\hat{\mathbf x}^{\rm T}\mathcal S \hat{\mathbf x},\label{eq.def_deltaS}
\end{eqnarray}
where $\mathcal S$ is a positive definite symmetric matrix and $\rm T$ denotes the transpose.
Note that we may regard $\delta^2 S$ as the total entropy change of the system and bath associated with the spontaneous fluctuation.
In general, in the vicinity of the equilibrium state, the dynamics of the fluctuation $\hat{\mathbf x}$ obeys the following Langevin equation~\cite{YO,LL}:
\begin{eqnarray}
\frac{d\hat{\mathbf x}}{dt}=L\frac{\partial \delta^2 S(\hat{\mathbf x})}{\partial \hat{\mathbf x}}+\bm \xi, \label{eq.LD}
\end{eqnarray}
where $L$ is Onsager's kinetic coefficients.
Onsager's kinetic coefficients $L$ are symmetric as $L=L^{\rm T}$, assuming that $\hat{\mathbf x}$ is the time-reversely symmetric quantities 
under time-reversely symmetric microscopic dynamics.
Moreover, $L$ is a positive definite matrix, as shown below.
$\bm \xi$ is Gaussian white noise satisfying $\left<\bm \xi(t) \right>=\mathbf 0$ and $\left<\bm \xi(t)\bm \xi(t')^{\rm T} \right>=2Lk_{\mathrm B}\delta(t-t')$, 
which assures that the stationary probability distribution of $\hat{\mathbf x}$ agrees with the Einstein's fluctuation formula eq.~(\ref{eq.Einstein}) (the fluctuation-dissipation theorem).

By defining thermodynamic forces $\hat{\mathbf y}$ as restoring forces to the equilibrium as
\begin{eqnarray}
\hat{\mathbf y}\equiv -\frac{\partial \delta^2 S(\hat{\mathbf x})}{\partial \hat{\mathbf x}}=\mathcal S\hat{\mathbf x},\label{eq.thermo_force}
\end{eqnarray}
equation~(\ref{eq.LD}) may be expressed as the linear flux-force relation $d\hat{\mathbf x}/dt=-L\hat{\mathbf y}+\bm \xi$.
By defining an ensemble average $\mathbf x\equiv \left<\hat{\mathbf x}\right>$ and $\mathbf y\equiv \left<\hat{\mathbf y}\right>$, 
the positive definiteness of $L$ is concluded from the positivity of $d\delta^2 S(\mathbf x)/dt$ during relaxation to the equilibrium~\cite{LL}: $d\delta^2 S(\mathbf x)/dt=-(\mathcal S\mathbf x)^{\rm T}\dot{\mathbf x}=\mathbf y^{\rm T}L\mathbf y \ge 0$, where we used $\dot{\mathbf x}=-L\mathbf y$ and the equality holds for $\mathbf y=\bf 0$.

By defining a relaxation matrix $A$ as
\begin{eqnarray}
A\equiv L\mathcal S,\label{eq.A}
\end{eqnarray}
which is a positive definite matrix reflecting the stability of the equilibrium state, we can write eq.~(\ref{eq.LD}) as
\begin{eqnarray}
\frac{d\hat{\mathbf x}}{dt}=-A\hat{\mathbf x}+\bm \xi. \label{eq.LD2}
\end{eqnarray}
We add small external thermodynamic forces $\mathbf F(t/\mathcal T_F)$ changing slowly with time from $t=0$ to $t=\mathcal T_F$ ($0 \le t \le \mathcal T_F$) to the system, which are intensive quantities as may be constituted with variations of intensive thermodynamic variables of the bath.
Then, the dynamics eq.~(\ref{eq.LD}) may be altered as~\cite{LL}
\begin{eqnarray}
\frac{d\hat{\mathbf x}}{dt}=L\frac{\partial}{\partial \hat{\mathbf x}}\left(\delta^2 S(\hat{\mathbf x})+\hat{\mathbf x}^{\rm T}\mathbf F(t/\mathcal T_F)\right)+\bm \xi, \label{eq.LD_F}
\end{eqnarray}
or, equivalently, 
\begin{eqnarray}
\frac{d\hat{\mathbf x}}{dt}=-L(\hat{\mathbf y}-\mathbf F(t/\mathcal T_F))+\bm \xi,\label{eq.LD_y}
\end{eqnarray}
using $\hat{\mathbf y}$, with $\hat{\mathbf y}-\mathbf F$ being considered as the effective thermodynamic forces.
Using Eq~(\ref{eq.A}), we can rewrite eq.~(\ref{eq.LD_F}) as
\begin{eqnarray}
\frac{d\hat{\mathbf x}}{dt}=-A\hat{\mathbf x}+L\mathbf F(t/\mathcal T_F)+\bm \xi.\label{eq.LD_F2}
\end{eqnarray}
By taking an ensemble average of both sides, we obtain the following:
\begin{eqnarray}
\frac{d \mathbf x}{dt}=-A\mathbf x+L\mathbf F(t/\mathcal T_F).\label{eq.LD_F3}
\end{eqnarray}
Or, equivalently, by taking an ensemble average of both sides of eq.~(\ref{eq.LD_y}), we obtain the following:
\begin{eqnarray}
\frac{d\mathbf x}{dt}=-L(\mathbf y-\mathbf F(t/\mathcal T_F)).\label{eq.LD_y2}
\end{eqnarray}

\section{Non-quasistatic response coefficients}\label{Non-quasistatic response coefficients}
Equation~(\ref{eq.LD_F3}) can be solved perturbatively using a two-timing method~\cite{Stz}.
We introduce a small dimensionless parameter $\epsilon \equiv t_s/\mathcal T_F \ll 1$ 
defined as the ratio of the typical time scale characterizing the relaxation of the system $t_s$ to the duration $\mathcal T_F$ required for a process changing $\mathbf F(t/\mathcal T_F)$ ($0 \le t \le \mathcal T_F$).
By introducing the dimensionless time $\tilde t\equiv t/t_s$ ($0 \le \tilde t \le 1/\epsilon$), we rewrite eq.~(\ref{eq.LD_F3}) as
\begin{eqnarray}
\frac{d \mathbf x}{d\tilde t}=-t_sA\mathbf x+t_s L\mathbf F(\epsilon \tilde t).\label{eq.LD_F3_2}
\end{eqnarray}
We then expand $\mathbf x(\tilde t)$ in terms of the fast and slow time scales $\tau \equiv \tilde t$ and $\mathcal T\equiv \epsilon \tilde t$ as
$\mathbf x(\tilde t,\epsilon)=\mathbf x^{(0)}(\tau,\mathcal T)+\epsilon \mathbf x^{(1)}(\tau,\mathcal T)+O(\epsilon^2)$.
The time-differential operator thus becomes $d\mathbf x/d\tilde t=\partial \mathbf x/\partial \tau+\epsilon \partial \mathbf x/\partial \mathcal T$.
By putting this into eq.~(\ref{eq.LD_F3_2}), the following equation for each order of $\epsilon$ is obtained:
\begin{eqnarray}
&&O(1):\frac{\partial \mathbf x^{(0)}}{\partial \tau}=-t_sA\mathbf x^{(0)}+t_s L\mathbf F(\mathcal T),\label{eq.x_0}\\
&&O(\epsilon):\frac{\partial \mathbf x^{(1)}}{\partial \tau}=-t_s A\mathbf x^{(1)}-\frac{\partial \mathbf x^{(0)}}{\partial \mathcal{T}}\label{eq.x_1}.
\end{eqnarray}
For each order, we consider a stationary solution with respect to the fast time, $\partial_\tau \mathbf x_{\rm s}^{(0)}=\partial_\tau \mathbf x_{\rm s}^{(1)}=0$, under the assumption of time-scale separation:
\begin{eqnarray}
\mathbf x_{\rm s}^{(0)}
=A^{-1}L\mathbf F(\mathcal T)=\mathcal S^{-1}\mathbf F(\mathcal T),\label{eq.xi_ad}
\end{eqnarray}
and
\begin{eqnarray}
\mathbf x_{\rm s}^{(1)}
=-t_s^{-1}A^{-1}\frac{\partial \mathbf x_{\rm s}^{(0)}}{\partial \mathcal T},\label{eq.xi_nad}
\end{eqnarray}
respectively. Note that the stationary $\mathbf x_{\rm s}^{(0)}$ and $\mathbf x_{\rm s}^{(1)}$ are dependent only on the slow time scale $\mathcal T$.
Consequently, we obtain $\mathbf x_{\rm s}=\mathbf x_{\rm s}^{(0)}+\epsilon \mathbf x_{\rm s}^{(1)}$ as
\begin{eqnarray}
\mathbf x_{\rm s}=\chi \mathbf F-\mathcal T_F^{-1}R \mathbf F'(\mathcal T)=\chi \mathbf F-R\dot{\mathbf F},\label{eq.main}
\end{eqnarray}
where the prime denotes the derivative with respect to $\mathcal T$, and the quasistatic response coefficients $\chi$ and the non-quasistatic response coefficients $R$ are given as
\begin{eqnarray}
&\chi=\mathcal S^{-1},\\
&R=A^{-1} \mathcal S^{-1}=\mathcal S^{-1}L^{-1}\mathcal S^{-1},\label{eq.R}
\end{eqnarray}
respectively.
It is remarkable that the non-quasistatic response coefficient $R$ is directly related to Onsager's kinetic coefficients $L$ that govern the relaxation dynamics of fluctuations of thermodynamic variables. 

Several remarks are in order with respect to the results eqs.~(\ref{eq.main})--(\ref{eq.R}).

From eq.~(\ref{eq.R}), it can be concluded that $R$ is symmetric as $R=R^{\rm T}$, using the symmetric Onsager's kinetic coefficients $L$ and the symmetric $\mathcal S$.
Moreover, $R$ is a positive definite matrix. This is confirmed by noticing that $R$ in eq.~(\ref{eq.R}) can be written as $R=(\mathcal S^{-1})^{\rm T}L^{-1}\mathcal S^{-1}$, where we used $\mathcal S^{-1}=(\mathcal S^{-1})^{\rm T}$. This shows that $L^{-1}$ and $R$ are congruent, and because $L^{-1}$ is positive definite, $R$ is also positive definite.

Equations~(\ref{eq.main})--(\ref{eq.R}) are consistent with the linear response theory.
We can show $\left<\hat{\mathbf x}\hat{\mathbf x}^{\rm T}\right>_{\rm eq}=k_{\mathrm B}\mathcal S^{-1}$, which implies that $\chi$ is calculated using 
the equilibrium correlation function of $\hat{\mathbf x}$ as $\chi=\left<\hat{\mathbf x}\hat{\mathbf x}^{\rm T}\right>_{\rm eq}/k_{\mathrm B}$, where $\left<\cdot \right>_{\rm eq}$ is taken with respect to $\mathcal P_{\rm eq}(\hat{\mathbf x})$ in eq.~(\ref{eq.Einstein}).
Moreover, in the linear response theory, $R$ can also be calculated using the time integral of the temporal equilibrium correlation function under an assumption of time-scale separation~\cite{SC2012,PP2021}:
\begin{eqnarray}
R=\frac{1}{k_{\mathrm B}}\int_0^\infty \left<\hat{\mathbf x}(t)\hat{\mathbf x}(0)^{\rm T}\right>dt.\label{eq.R_CR}
\end{eqnarray}
This can be easily shown by substituting the explicit solution 
\begin{eqnarray}
\hat{\mathbf x}(t)=e^{-At}\hat{\mathbf x}(0)+\int_0^t e^{-A(t-t')}\bm \xi(t')dt'
\end{eqnarray}
of eq.~(\ref{eq.LD_F2})
into eq.~(\ref{eq.R_CR}).
Thus, eq.~(\ref{eq.R}) shows the detailed constituents of $R$ for the fluctuations of thermodynamic variables governed by eq.~(\ref{eq.LD}).

\section{Thermodynamic length and dissipated availability}\label{Dissipated availability}
We introduce the dissipated availability $A_{\rm diss}$, which quantifies the efficiency of irreversible thermodynamic processes~\cite{SB1983}.
Let us consider the following non-negative quantity as $A_{\rm diss}$ using the time derivative of the second-order entropy variation with its quasistatic value being subtracted (see appendix for the derivation):
\begin{eqnarray}
A_{\rm diss}&\equiv \int_0^{\mathcal T_F} \frac{d}{dt}\Biggl [\delta^2 S(\mathbf x)-\left(-\frac{1}{2}{\mathbf x_{\rm s}^{(0)}}^{\rm T}\mathcal S\mathbf x_{\rm s}^{(0)}\right)\Biggr ]dt\nonumber\\
&\simeq \int_0^{\mathcal T_F} \dot{\mathbf F}^{\rm T}R\dot{\mathbf F}dt \ge 0,\label{eq.A_diss}
\end{eqnarray}
where the inequality holds by using the positive definiteness of $R$ shown above. 
As will be shown below, this quantity agrees with the total entropy production during $\mathcal T_F$.

By using the Cauchy-Schwartz inequality, we can obtain the tighter bound for $A_{\rm diss}$ than eq.~(\ref{eq.A_diss}):
\begin{eqnarray}
A_{\rm diss}\ge \frac{\mathcal L^2}{\mathcal T_F}\equiv A_{\rm diss}^*,\label{eq.A_diss_length}
\end{eqnarray}
where
\begin{eqnarray}
\mathcal L\equiv \int_{\gamma_F} \sqrt{d\mathbf F^{\rm T}Rd\mathbf F}=\int_0^1 \sqrt{{\mathbf F'}^{\rm T}R\mathbf F'}d\mathcal T\label{eq.length}
\end{eqnarray}
is the thermodynamic length for the path $\gamma_F$ on the control space of thermodynamics forces $\mathbf F$ with $R$ the metric tensor defined on it~\cite{SB1983,BS2020}.
The equality of eq.~(\ref{eq.A_diss_length}), the minimum dissipated availability $A_{\rm diss}^*$, is achieved for a geodesic path that yields the constant dissipation such that the integrand of eq.~(\ref{eq.length}) becomes constant~\cite{SB1983}.
As $R$ is the constant matrix evaluated at $\mathbf X_{\rm eq}$, this equality is expected to be realized for $\mathbf F$ with a linear profile.
From a geometric perspective, the space in the vicinity of the equilibrium state with the constant metric $R$ can be considered as a flat space.

It is also noteworthy that $A_{\rm diss}$ in eq.~(\ref{eq.A_diss}) can also be expressed in terms of $\dot{\mathbf x}_{\rm s}$ instead of $\dot{\mathbf F}$:
\begin{eqnarray}
A_{\rm diss}=\int_0^{\mathcal T_F} (\dot{\mathbf x}_{\rm s}^{(0)})^{\rm T}L^{-1}\dot{\mathbf x}_{\rm s}^{(0)} dt\ge 0, \label{eq.dissi_func}
\end{eqnarray}
where we used $\dot{\mathbf F}=\mathcal S\dot{\mathbf x}_{\rm s}^{(0)}$ derived from the time derivative of eq.~(\ref{eq.xi_ad}) and eq.~(\ref{eq.R}). 
The integrand of eq.~(\ref{eq.dissi_func}) is essentially the same quantity as the dissipation function~\cite{SS1997}.
The inequality in eq.~(\ref{eq.dissi_func}) is assured by the positive definiteness of $L^{-1}$.

Moreover, when expressed in terms of the effective thermodynamic force $\mathbf y-\mathbf F$, we obtain
\begin{eqnarray}
A_{\rm diss}=\int_0^{\mathcal T_F} ({\mathbf y}_{\rm s}-\mathbf F)^{\rm T}L(\mathbf y_{\rm s}-\mathbf F)dt\ge 0, \label{eq.entropy_product}
\end{eqnarray}
where $\mathbf y_{\rm s}=\mathcal S\mathbf x_{\rm s}$ and we used eq.~(\ref{eq.LD_y2}) to eq.~(\ref{eq.dissi_func}). The integrand of eq.~(\ref{eq.entropy_product}) takes the form of  
a ``familiar" total entropy production rate.
The inequality in eq.~(\ref{eq.entropy_product}) is assured by the positive definiteness of $L$.
These apparently different but the equivalent expressions eqs.~(\ref{eq.A_diss}), (\ref{eq.dissi_func}), and (\ref{eq.entropy_product}) unified in terms of the Onsager's kinetic coefficients are informative, 
which will be used for numerical demonstration of the present theory.
Note that as what is controlling here is the intensive thermodynamic force $\mathbf F$, the expression eq.~(\ref{eq.A_diss}) using $\dot{\mathbf F}$ may serve as the most natural form.
It should be noted that eq.~(\ref{eq.dissi_func}) using the extensive variables as the control parameters is approximated as
\begin{eqnarray}
A_{\rm diss}=\int_0^{\mathcal T_F} (\dot{\mathbf x}_{\rm s}^{(0)})^{\rm T}\mathcal S \tau_c \dot{\mathbf x}_{\rm s} dt, \label{eq.dissi_func_2}
\end{eqnarray}
where we used $L^{-1}=\mathcal SA^{-1}$ and further replaced $A^{-1}$, with its components being having a dimension of time, with a correlation (relaxation) time matrix $\tau_c$~\cite{SC2012,KA1997,BD2014} (or a lag time~\cite{SB1983,FAQS1985,TSM1994}). 
Similar but not exactly the same expressions to eq.~(\ref{eq.dissi_func_2}) have been obtained in stochastic thermodynamic systems~\cite{SC2012,KA1997,BD2014} and in endoreversible systems~\cite{SB1983,FAQS1985,TSM1994}.
In the latter case, the metric tensor $\mathcal S$ in eq.~(\ref{eq.dissi_func_2}) is given in the entropic form~\cite{R1979,SNI1984} rather than the energetic form~\cite{SB1983}.
Therefore, our formulation using Onsager's kinetic coefficients may provide a solid basis for the endoreversible formulation in~\cite {SB1983}.

We can replace some of the extensive variables in eq.~(\ref{eq.dissi_func}) with intensive variables by a variable transformation from $\mathbf X$ to $\mathbf Z$~\cite{SNI1984}:
\begin{eqnarray}
A_{\rm diss}=\int_0^{\mathcal T_F} {\dot{\mathbf z}}^{\rm T} \eta \dot{\mathbf z}dt\ge 0, \label{eq.dissi_func_Legendre}
\end{eqnarray}
where 
\begin{eqnarray}
\eta \equiv Q^{\rm T}L^{-1}Q,
\end{eqnarray}
and $Q$ is a Jacobian matrix:
\begin{eqnarray}
Q\equiv \frac{\partial (X_1, \cdots, X_n)}{\partial (Z_1, \cdots, Z_n)}.
\end{eqnarray}
For example, when the internal energy $U$ is transformed to the temperature $T$, the Jacobian matrix should be~\cite{SNI1984}
\begin{eqnarray}
Q=\frac{\partial (U, V)}{\partial (T, V)}
=\left(
\begin{array}{cc}
(\partial U/\partial T)_V & (\partial U/\partial V)_T\\
0 & 1
\end{array}
\right).
\end{eqnarray}

\begin{figure}[!t]
\begin{center}
\includegraphics[scale=0.6]{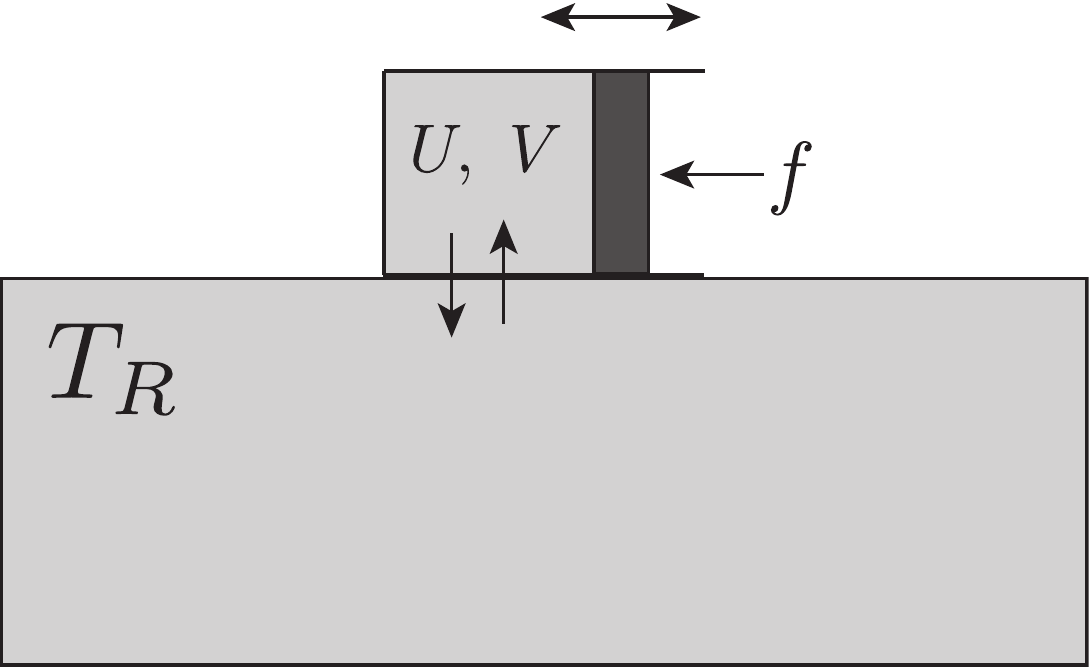}
\caption{Schematic illustration of a thermodynamic system under consideration: The ideal gas is confined in the cylinder enclosed by the walls
with a movable piston on one side to which the external mechanical force $f$ is applied. Further, it is in thermal contact with a heat bath at changeable temperature $T_R$, which may be realized, {\it e.g.,} by putting multiple heat baths differentiated with continuously changing temperature labels~\cite{YI2022}.}\label{setup}
\end{center}
\end{figure}

\section{Example}\label{Example}
As a demonstration of our theory, we identify the non-quasistatic response coefficients of a macroscopic phenomenological model of a thermodynamic system.
Our model is an ideal gas confined in a cylinder with a movable piston subject to an external mechanical force $f(t/\mathcal T_F)$ in contact with a heat bath at changeable temperature $T_R(t/\mathcal T_F)$ (fig.~\ref{setup}).
Let $\mathbf X=(U, V)$ ($n=2$) be the extensive thermodynamic variables of the ideal gas, where $U=C_VT=f_dNk_{\mathrm B}T/2$ is the internal energy with the constant-volume heat capacity $C_V$, 
the temperature of the gas $T$, and the internal degrees of freedom $f_d$. Further, $V$ is the volume of the ideal gas.
Under an assumption of spatial uniformity (or endoreversibility~\cite{SB1983}), macroscopic dynamics for $\mathbf X$ is given by
\begin{eqnarray}
&\frac{dU}{dt}=\kappa (T_R(t/\mathcal T_F)-T)-\frac{f(t/\mathcal T_F)}{\sigma}\frac{dV}{dt},\label{eq.1st_law}\\
&\frac{dV}{dt}=\frac{\sigma^2}{\Gamma}\left(\frac{Nk_{\mathrm B}T}{V}-\frac{f(t/\mathcal T_F)}{\sigma}\right).\label{eq.piston}
\end{eqnarray}
Equation~(\ref{eq.1st_law}) is the first law of thermodynamics (the energy conservation law) per unit time, where the first term on the right-hand side denotes the Fourier's law for the heat flow with $\kappa$ the thermal conductance
and the second term does the work flow done by the gas against the mechanical force $f(t/\mathcal T_F)$ acting on the piston in one direction, where $\sigma$ is the section area of the piston.
Equation~(\ref{eq.piston}) is the equation of motion for an overdamped piston whose inertia can be neglected, where $\Gamma$ is the friction coefficient of the piston.
The piston moves back and forth subject to the net force due to the internal gas pressure $Nk_{\mathrm B}T/V$ and the mechanical pressure $f(t/\mathcal T_F)/\sigma$.

For the fixed temperature and mechanical force $T_R(t/\mathcal T_F)=T_{\rm eq}$ and $f(t/\mathcal T_F)=f_{\rm eq}$, respectively,
the stationary state $\mathbf X_{\rm eq}=(U_{\rm eq}, V_{\rm eq})=(C_VT_{\rm eq}, \sigma Nk_{\mathrm B}T_{\rm eq}/f_{\rm eq})$ satisfying $d\mathbf X/dt=\bm 0$
corresponds to the equilibrium state. 

Equations~(\ref{eq.1st_law}) and (\ref{eq.piston}) around the equilibrium state $\mathbf X_{\rm eq}$ can be rewritten into the form eq.~(\ref{eq.LD_F3}).
First, consider the case of fixed temperature and mechanical force $T_R(t/\mathcal T_F)=T_{\rm eq}$ and $f(t/\mathcal T_F)=f_{\rm eq}$.
By expanding eqs.~(\ref{eq.1st_law}) and (\ref{eq.piston}) in terms of $\mathbf x$ around $\mathbf X_{\rm eq}$, $d\mathbf x/dt=-A\mathbf x$ is obtained, neglecting the higher order terms of $\mathbf x$, where $A$ is given by
\begin{eqnarray}
A
=
\left(
\begin{array}{cc}
\frac{\kappa}{C_V}+f_{\rm eq}\frac{\sigma}{\Gamma}\frac{Nk_{\mathrm B}}{C_VV_{\rm eq}} & -f_{\rm eq}\frac{\sigma}{\Gamma}\frac{Nk_{\mathrm B}T_{\rm eq}}{V_{\rm eq}^2} \\
-\frac{\sigma^2}{\Gamma}\frac{Nk_{\mathrm B}}{C_VV_{\rm eq}} & \frac{\sigma^2}{\Gamma}\frac{Nk_{\mathrm B}T_{\rm eq}}{V_{\rm eq}^2}
\end{array}
\right).\label{eq.A_relx}
\end{eqnarray}
The Einstein's fluctuation formula for the corresponding fluctuation $\hat{\mathbf x}=\delta \hat{\mathbf X}=(\delta \hat U, \delta \hat V)$ is given as~\cite{LL}
\begin{eqnarray}
\delta^2 S(\hat{\mathbf x})=-\frac{1}{2}\left[\frac{1}{C_VT_{\rm eq}^2}\delta \hat U^2+\frac{Nk_{\mathrm B}}{V_{\rm eq}^2}\delta \hat V^2\right],
\end{eqnarray}
from which $\mathcal S_{11}=\frac{1}{C_VT_{\rm eq}^2}$, $\mathcal S_{12}=\mathcal S_{21}=0$, and $\mathcal S_{22}=\frac{Nk_{\mathrm B}}{V_{\rm eq}^2}$ are identified.
Then, we obtain the quasistatic response coefficient $\chi=\mathcal S^{-1}$ as
\begin{eqnarray}
\chi
=
\left(
\begin{array}{cc}
C_VT_{\rm eq}^2 & 0 \\
0 & \frac{V_{\rm eq}^2}{Nk_{\mathrm B}}
\end{array}
\right).\label{eq.chi_gas}
\end{eqnarray}
We can also obtain the Onsager's kinetic coefficients $L=A\mathcal S^{-1}$ as
\begin{eqnarray}
L=
\left(
\begin{array}{cc}
\kappa T_{\rm eq}^2+\frac{f_{\rm eq}^2}{\Gamma}T_{\rm eq} & -f_{\rm eq}\frac{\sigma}{\Gamma}T_{\rm eq} \\
 -f_{\rm eq}\frac{\sigma}{\Gamma}T_{\rm eq} & \frac{\sigma^2}{\Gamma}T_{\rm eq}
\end{array}
\right).\label{eq.L}
\end{eqnarray}
Next, we consider the effect of time-dependent temperature of the heat bath $T_R(t/\mathcal T_F)=T_{\rm eq}+\delta T_R(t/\mathcal T_F)$ and mechanical force $f(t/\mathcal T_F)=f_{\rm eq}+\delta f(t/\mathcal T_F)$, where $|\delta T_R(t/\mathcal T_F)| \ll T_{\rm eq}$ and $|\delta f(t/\mathcal T_F)| \ll f_{\rm eq}$.
By expanding eqs.~(\ref{eq.1st_law}) and (\ref{eq.piston}) up to $O(\mathbf x, \delta T_R, \delta f)$, we obtain
\begin{eqnarray}
\frac{d\mathbf x}{dt}=-A\mathbf x+
\left(
\begin{array}{cc}
\kappa \delta T_R(t/\mathcal T_F)+\frac{f_{\rm eq}}{\Gamma}\delta f(t/\mathcal T_F) \\
 -\frac{\sigma}{\Gamma}\delta f(t/\mathcal T_F)
\end{array}
\right).\label{eq.expand}
\end{eqnarray}
Through comparisons of eq.~(\ref{eq.expand}) and eq.~(\ref{eq.LD_F3}),
the external thermodynamic force $\mathbf F(t/\mathcal T_F)$ acting on $\mathbf x$ is identified as
\begin{eqnarray}
\mathbf F(t/\mathcal T_F)
=\left(
\begin{array}{cc}
\frac{\delta T_R(t/\mathcal T_F)}{T_{\rm eq}^2} \\
\frac{\delta p_R(t/\mathcal T_F)}{T_{\rm eq}}-\frac{\delta f(t/\mathcal T_F)/\sigma}{T_{\rm eq}}
\end{array}
\right),
\end{eqnarray}
where $\delta p_R(t/\mathcal T_F)\equiv Nk_{\mathrm B}\delta T_R(t/\mathcal T_F)/V_{\rm eq}$.
Here, $F_2(t/\mathcal T_F)$ denotes the net force increment acting on the piston that arises owing to the combined effect of the change of the gas pressure upon the temperature change of the heat bath and the change of the external mechanical force.
Using eqs.~(\ref{eq.R}), (\ref{eq.chi_gas}), and (\ref{eq.L}), the non-quasistatic response coefficient $R$ is derived as
\begin{eqnarray}
R=\left(
\begin{array}{cc}
\frac{C_V^2T_{\rm eq}^2}{\kappa} & \frac{C_V T_{\rm eq}V_{\rm eq}}{\kappa} \\
 \frac{C_V T_{\rm eq}V_{\rm eq}}{\kappa} & \frac{V_{\rm eq}^2}{\kappa}+\frac{f_d^2\Gamma V_{\rm eq}^4}{4\sigma^2C_V^2T_{\rm eq}}
\end{array}
\right).\label{eq.R_gas}
\end{eqnarray}
Through explicit writing, the following is obtained as eq.~(\ref{eq.main}):
\begin{eqnarray}
&\delta U_{\rm s}(\mathcal T)=\chi_{11}F_1(\mathcal T)-\mathcal T_F^{-1} R_{11}F'_1(\mathcal T)-\mathcal T_F^{-1} R_{12}F'_2(\mathcal T),\label{eq.deltaU}\\
&\delta V_{\rm s}(\mathcal T)=\chi_{22}F_2(\mathcal T)-\mathcal T_F^{-1} R_{21}F'_1(\mathcal T)-\mathcal T_F^{-1} R_{22}F'_2(\mathcal T).\label{eq.deltaV}
\end{eqnarray}
Note that while the non-diagonal components of $\chi$ in eq.~(\ref{eq.chi_gas}) identically vanish,
the non-diagonal components of $R$ in eq.~(\ref{eq.R_gas}) are non-zero. Therefore, the cross effect between $\delta U$ and $\delta V$ appears as a consequence of a nonequilibrium control of the external thermodynamic forces.

\begin{table}
\caption{\label{table}Summary for nondimensionalized quantities}
\begin{indented}
\item[]
\begin{tabular}{lcr}
\hline
\\ [-1.5ex]
Time $t$ & $\tilde t=t/t_s$ with $t_s=C_V/\kappa$ \\
Internal energy $U$ & $\tilde U=U/U_{\rm eq}=U/(C_VT_{\rm eq})$ \\
Volume $V$ & $\tilde V=V/V_{\rm eq}$\\
Temperature $T$, $T_R$ & $\tilde T=T/T_{\rm eq}$, $\tilde T_R=T_R/T_{\rm eq}$\\
Friction coefficient $\Gamma$ & $\tilde \Gamma=\Gamma/(\sigma^2 C_V^2 T_{\rm eq}/\kappa V_{\rm eq}^2)$\\
Mechanical force $f$ & $\tilde f=f/(\sigma C_V T_{\rm eq}/V_{\rm eq})$\\ [1ex]
\hline
\end{tabular}
\end{indented}
\end{table}

\begin{table}
\caption{\label{table2}Summary for nondimensionalized $\mathbf F$, $\chi$, and $R$}
\begin{indented}
\item[]
\begin{tabular}{lcr}
\hline 
\\ [-1.5ex]
$F_1(\mathcal T)$ & $\tilde F_1(\mathcal T)=T_{\rm eq}F_1(\mathcal T)=\delta \tilde{T}_R(\mathcal T)$\\
$F_2(\mathcal T)$ & $\tilde F_2(\mathcal T)=\frac{V_{\rm eq}}{C_V}F_2(\mathcal T)=2\delta \tilde{T}_R(\mathcal T)/f_d-\delta \tilde{f}(\mathcal T)$\\
$\chi_{11}$ & $\tilde \chi_{11}=1$\\
$\chi_{22}$ & $\tilde \chi_{22}=f_d/2$\\
$R_{11}$ & $\tilde R_{11}=1$ \\
$R_{12}$ & $\tilde R_{12}=1$\\ 
$R_{21}$ & $\tilde R_{21}=1$ \\
$R_{22}$ & $\tilde R_{22}=1+f_d^2 \tilde \Gamma/4$\\ [1ex]
\hline
\end{tabular}
\end{indented}
\end{table}

\section{Numerical demonstration}\label{Numerical demonstration}
We numerically confirm the validity of eqs.~(\ref{eq.deltaU}) and (\ref{eq.deltaV}).
For this purpose, eqs.~(\ref{eq.1st_law}) and (\ref{eq.piston}) are made nondimensional as follows:
\begin{eqnarray}
&\frac{d\tilde U}{d\tilde t}=\tilde T_R(\epsilon \tilde t)-\tilde U-\tilde f(\epsilon \tilde t)\frac{d\tilde V}{d\tilde t},\label{eq.1st_law_nondim}\\
&\frac{d\tilde V}{d\tilde t}=\frac{1}{\tilde \Gamma}\left(\frac{2\tilde U}{f_d\tilde V}-\tilde f(\epsilon \tilde t)\right),\label{eq.piston_nondim}
\end{eqnarray}
respectively, where we put $T=U/C_V$ on the right-hand sides of eqs.~(\ref{eq.1st_law}) and (\ref{eq.piston}) before the nondimensionalization 
and the nondimensionalized quantities are summarized in Table~\ref{table}.
The quantities with a tilde denote nondimensionalized quantities.
Then, eqs.~(\ref{eq.deltaU}) and (\ref{eq.deltaV}) are also nondimensionalized as
\begin{eqnarray}
&\delta \tilde U_{\rm s}(\mathcal T)=\tilde \chi_{11}\tilde F_1(\mathcal T)-\epsilon \tilde R_{11}{\tilde F}'_1(\mathcal T)-\epsilon \tilde{R}_{12}{\tilde F}'_2(\mathcal T),\label{eq.deltaU_nondim}\\
&\delta \tilde V_{\rm s}(\mathcal T)=\tilde \chi_{22}\tilde F_2(\mathcal T)-\epsilon \tilde R_{21}{\tilde F}'_1(\mathcal T)-\epsilon \tilde{R}_{22}{\tilde F}'_2(\mathcal T),\label{eq.deltaV_nondim}
\end{eqnarray}
in terms of $\epsilon$, respectively, where the nondimensionalized $\mathbf F$, $\chi$, and $R$ are summarized in Table~\ref{table2}.

\begin{figure}[!t]
\begin{center}
\includegraphics[scale=1.2]{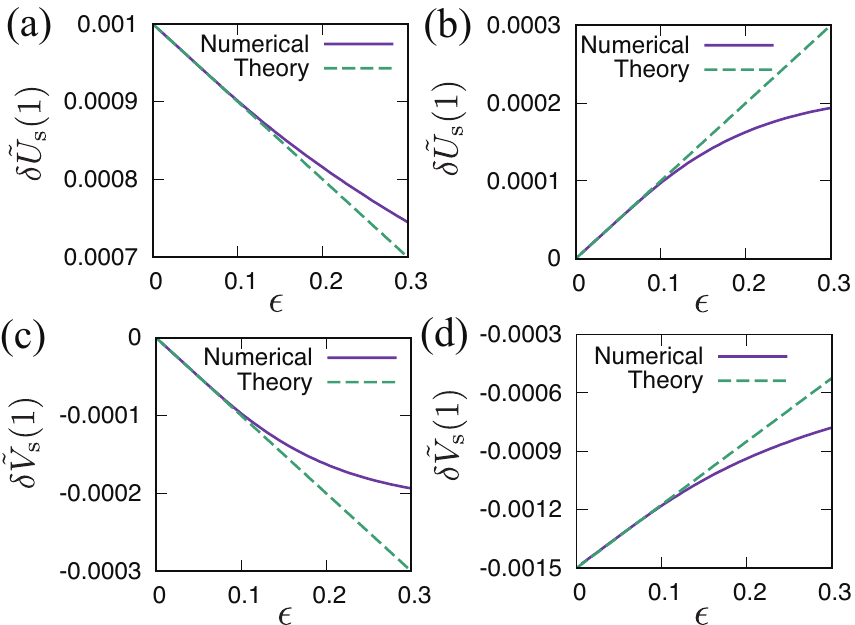}
\caption{(a) and (b): $\epsilon$ dependence of $\delta \tilde U_{\rm s}(1)$ for (a) $\tilde F_2'(1)=0$ and (b) $\tilde F_1'(1)=0$. 
The theory denotes eq.~(\ref{eq.deltaU_nondim}) with $\mathcal T=1$.
(c) and (d): $\epsilon$ dependence of $\delta \tilde V_{\rm s}(1)$ for (c) $\tilde F_2'(1)=0$ and (d) $\tilde F_1'(1)=0$. The theory denotes eq.~(\ref{eq.deltaV_nondim}) with $\mathcal T=1$. 
As parameter values, $f_d=3$, $\tilde \Gamma=1$, $\Delta \tilde T=10^{-3}$, and $\Delta \tilde f=10^{-3}$ were used. The fourth Runge--Kutta method with time step $10^{-3}$ was used for obtaining the numerical results.}\label{fig2}
\end{center}
\end{figure}

In fig.~\ref{fig2}, the theoretical results eqs.~(\ref{eq.deltaU_nondim}) and (\ref{eq.deltaV_nondim}) for $\mathcal T=1$ are compared with the numerical results obtained by solving eqs.~(\ref{eq.1st_law_nondim}) and (\ref{eq.piston_nondim}) numerically with the following linear profile as $\delta \tilde{T}_R(\mathcal T)$ and $\delta \tilde{f}(\mathcal T)$ ($0 \le \mathcal T \le 1$):
\begin{eqnarray}
\delta \tilde{T}_R(\mathcal T)=\Delta \tilde T \mathcal T, \ \delta \tilde{f}(\mathcal T)=\Delta \tilde f \mathcal T, \label{eq.linear_protocol}
\end{eqnarray}
where $0 \le \Delta \tilde{T}\ll 1$ and $0 \le \Delta \tilde{f}\ll 1$ are small increments.
As evident, for sufficiently small $\epsilon$, the numerical results are consistent with the theoretical results as expected, with the discrepancies observed with an increase in $\epsilon$ where
the nonlinear effects begin to appear.

\begin{figure}[!t]
\begin{center}
\includegraphics[scale=1.5]{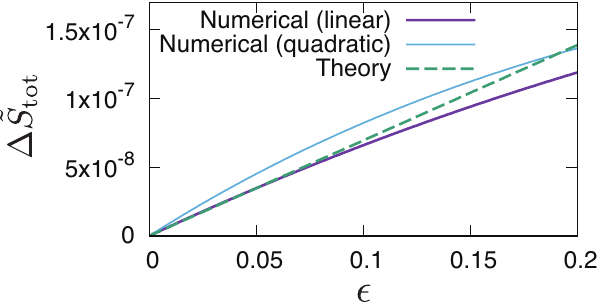}
\caption{$\epsilon$ dependence of the total entropy production $\Delta \tilde{S}_{\rm tot}$ in eq.~(\ref{eq.DS_tot_nondim}) calculated for eq.~(\ref{eq.linear_protocol}) (bold solid line) and for eq.~(\ref{eq.periodic_quadratic_profile}) (thin solid line).
The theory denotes the theoretical minimum dissipated availability $\tilde A_{\rm diss}^*$ in eq.~(\ref{eq.A_diss_nondim_linear}).
The same parameter values as those in fig.~\ref{fig2} and the fourth Runge--Kutta method with time step $10^{-4}$ were used.}\label{fig3}
\end{center}
\end{figure}

Finally, we demonstrate the dissipated availability $A_{\rm diss}$ in eq.~(\ref{eq.A_diss}). 
It is nondimensionalized as
\begin{eqnarray}
\tilde A_{\rm diss}^*=\frac{A_{\rm diss}^*}{C_V}=\epsilon \left(\int_0^1 \sqrt{{\tilde{\mathbf F}'^{\rm T}} \tilde R\tilde{\mathbf F}'} d\mathcal T\right)^2=\epsilon \tilde{\mathcal L}^2.\label{eq.A_diss_nondim}
\end{eqnarray}
As we know from eq.~(\ref{eq.entropy_product}) that the dissipated availability is equivalent to the total entropy production for small $\epsilon$, 
we introduce the total entropy production $\Delta S_{\rm tot}$ for comparison with $A_{\rm diss}$:
\begin{eqnarray}
\Delta S_{\rm tot}=\Delta S+\Delta S_{\rm bath},\label{eq.DeltaS_tot}
\end{eqnarray}
which is the sum of the entropy change of the system and that of the heat bath given as
\begin{eqnarray}
\Delta S&=S(T(\mathcal T_F), V(\mathcal T_F))-S(T(0), V(0))\nonumber\\
&=C_V \ln \frac{T(\mathcal T_F)}{T_{\rm eq}}+Nk_{\rm B}\ln \frac{V(\mathcal T_F)}{V_{\rm eq}},\\
\Delta S_{\rm bath}&=-\int_0^{\mathcal T_F} \frac{\kappa(T_R(t/\mathcal T_F)-T(t))}{T_R(t/\mathcal T_F)}dt\nonumber\\
&=-\kappa \int_0^{\mathcal T_F}\left(1-\frac{T(t)}{T_R(t/\mathcal T_F)}\right)dt.
\end{eqnarray}
Here, $S(T,V)\equiv C_V\ln T+Nk_{\rm B}\ln V+S_0$ is the thermodynamic entropy of the ideal gas, where $S_0$ is the value at a reference state.
The nondimensionalized form of eq.~(\ref{eq.DeltaS_tot}) reads
\begin{eqnarray}
\Delta \tilde{S}_{\rm tot}=\frac{\Delta S_{\rm tot}}{C_V}&=\ln \tilde T\left(1/\epsilon\right)+\frac{2}{f_d}\ln \tilde V\left(1/\epsilon \right)\nonumber\\
&-\int_0^{1/\epsilon} \left(1-\frac{\tilde T(\tilde t)}{\tilde T_R(\epsilon \tilde t)}\right)d\tilde t.\label{eq.DS_tot_nondim}
\end{eqnarray}
For comparison with the linear protocol in eq.~(\ref{eq.linear_protocol}) as the geodesic path, we also use $\mathbf F$ with the following quadratic protocol as the non-geodesic path $(0 \le \mathcal T \le 1)$:
\begin{eqnarray}
\delta \tilde{T}_R(\mathcal T)=\Delta \tilde T \mathcal T^2, \ \delta \tilde{f}(\mathcal T)=\Delta \tilde f \mathcal T^2.\label{eq.periodic_quadratic_profile}
\end{eqnarray}
As we noted, the minimum dissipated availability $\tilde A_{\rm diss}^*$ is expected to be achieved for the linear protocol in eq.~(\ref{eq.linear_protocol}). 
Explicitly, $\tilde A_{\rm diss}^*$ calculated for eq.~(\ref{eq.linear_protocol}) is given as
\begin{eqnarray}
\tilde A_{\rm diss}^*=\epsilon \Biggl \{\Delta \tilde{T}^2&+2\Delta \tilde{T}\left(\frac{2\Delta \tilde{T}}{f_d}-\Delta \tilde f\right)\nonumber\\
&+\left(1+\frac{f_d^2 \tilde \Gamma}{4}\right)\left(\frac{2\Delta \tilde{T}}{f_d}-\Delta \tilde f\right)^2\Biggr \}.\label{eq.A_diss_nondim_linear}
\end{eqnarray}

In fig.~\ref{fig3}, we show the total entropy production $\Delta \tilde{S}_{\rm tot}$ numerically calculated for the linear protocol eq.~(\ref{eq.linear_protocol}) and for the quadratic protocol eq.~(\ref{eq.periodic_quadratic_profile}),
with a comparison with the minimum dissipated availability $\tilde A_{\rm diss}^*$ in eq.~(\ref{eq.A_diss_nondim_linear}).
We can confirm that $\Delta \tilde{S}_{\rm tot}$ for the linear protocol as the geodesic path agrees with $\tilde A_{\rm diss}^*$ in the small $\epsilon$ region as expected.
Meanwhile, we can also confirm that $\Delta \tilde{S}_{\rm tot}$ for the quadratic protocol as the non-geodesic path is larger than $\tilde A_{\rm diss}^*$ in the small $\epsilon$ region, which is consistent with 
the theoretical prediction.

\section{Concluding perspective}\label{Concluding perspective}
We derived a simple formula for the non-quasistatic response coefficients for macroscopic thermodynamic systems.
The formula revealed the general relationship between the non-quasistatic response coefficients and Onsager's kinetic coefficients that govern the relaxation dynamics of fluctuations of semi-macroscopic thermodynamic variables.
Similarly to the quasistatic response coefficients whose symmetry has been already established in equilibrium thermodynamics,
the non-quasistatic response coefficients are also symmetric as $R=R^{\rm T}$, which is related to the symmetry of Onsager's kinetic coefficients.
Moreover, we also formulated the dissipated availability $A_{\rm diss}$ for the present system that quantifies the efficiency of irreversible thermodynamic processes.
It is expressed in terms of the time derivative of the second-order entropy variation, and the equivalent expressions using the dissipation function or the total entropy production rate were provided. Our theory was demonstrated by using the ideal gas model.

In this paper, we have considered only infinitesimally small changes of external forces in a macroscopic sense. 
The thermodynamic space is, therefore, necessarily flat with the constant metric. For extending the present theory to arbitrary changes between equilibrium states, 
it would be necessary that each equilibrium space is connected appropriately to form a globally curved space.

It is expected to measure the predicted symmetry of $R$ and the dissipated availability $A_{\rm diss}$ in realistic nonequilibrium experiments
such as those illustrated in fig.~\ref{setup} where eqs.~(\ref{eq.1st_law}) and (\ref{eq.piston}) serve as a good approximation.
Real systems may be accompanied by complicated fluid motion as well as the non-uniform heat conduction inside the system associated with a piston motion.
However, from recent experiments on real heat engines using a gas~\cite{MZCPD2020,TI2020}, it can be concluded that simple models similar to those presented in eqs.~(\ref{eq.1st_law}) and (\ref{eq.piston}) explain the experimental results to a sufficiently good approximation.
Therefore, experimental verification will be an important direction for future investigations.

Finally, finding a similar formula for linear response in nonequilibrium steady states~\cite{MPRV2008,BM2013,SS2010} will also be an important direction for future investigations.

\section*{Acknowledgments}
This work was supported by JSPS KAKENHI Grant Numbers 19K03651 and 22K03450.

\section*{Appendix. Derivation of eq.~(\ref{eq.A_diss})}\label{append_derivation_A_diss}
First, we evaluate the time derivative of the entropy variation $d\delta^2 S(\mathbf x)/dt=-d(\mathbf x^{\rm T}\mathcal S \mathbf x/2)/dt$:
\begin{eqnarray}
\frac{d\delta^2 S(\mathbf x)}{dt}&=-(\mathcal S\mathbf x)^{\rm T}\dot{\mathbf x}=-\mathbf y^{\rm T}\dot{\mathbf x}=\mathbf y^{\rm T}L(\mathbf y-\mathbf F),\label{eq.deltaS_append}
\end{eqnarray}
where we used eq.~(\ref{eq.LD_y2}) in the third equality. 
By applying $\mathcal S$ to both sides of eq.~(\ref{eq.main}), we can approximate $\mathbf y$ as
\begin{eqnarray}
\mathbf y\simeq {\mathbf y}_{\rm s} \equiv \mathcal S\mathbf x_{\rm s}=\mathbf F-\mathcal S R\dot{\mathbf F}.\label{eq.y_append}
\end{eqnarray}
By substituting eq.~(\ref{eq.y_append}) into eq.~(\ref{eq.deltaS_append}), we obtain
\begin{eqnarray}
\frac{d\delta^2 S(\mathbf x)}{dt}&\simeq -\mathbf F^{\rm T}\mathcal S^{-1}\dot{\mathbf F}+\dot{\mathbf F}^{\rm T}R\dot{\mathbf F}\nonumber\\
&=-\frac{d}{dt}\left(\frac{1}{2}\mathbf F^{\rm T}\mathcal S^{-1}\mathbf F\right)+\dot{\mathbf F}^{\rm T}R\dot{\mathbf F}\nonumber\\
&=-\frac{d}{dt}\left(\frac{1}{2}{\mathbf x_{\rm s}^{(0)}}^{\rm T}\mathcal S\mathbf x_{\rm s}^{(0)}\right)+\dot{\mathbf F}^{\rm T}R\dot{\mathbf F},\label{eq.deltaS_append_2}
\end{eqnarray}
from which we derive eq.~(\ref{eq.A_diss}).

\end{document}